\documentclass[11pt,a4paper]{article}
\pdfoutput=1
\usepackage{jheppub}
\usepackage[T1]{fontenc}
\usepackage{fix-cm}
\usepackage{lmodern}
\usepackage{amsmath,amssymb,mathtools,mathrsfs,empheq}
\usepackage{physics}
\usepackage{bbm}
\usepackage{bm}
\usepackage{hhline}
\usepackage{diagbox}
\usepackage{pdflscape}
\usepackage{fancyhdr} 

\fancypagestyle{mylandscape}{
\fancyhf{} 
\fancyfoot{
\makebox[\textwidth][r]{
  \rlap{\hspace{.75cm}
    \smash{
      \raisebox{4.87in}{
        \rotatebox{90}{\thepage}}}}}}
}

\makeatletter
\newcommand*{\rom}[1]{\expandafter\@slowromancap\romannumeral #1@}
\makeatother

\usepackage{tikz}

\usepackage{xcolor}

\newcommand\fft[2]{\frac{#1}{#2}}

\def\SO{{\rm SO}}

\def\U{{\rm U}}

\usepackage{tensor}

\title{Holographic Thermal Observables and M2-branes}
\author[a]{Nikolay Bobev,}
\author[a]{Junho Hong,}
\author[b]{and Valentin Reys}

\affiliation[a]{Instituut voor Theoretische Fysica, KU Leuven, \\
	Celestijnenlaan 200D, B-3001 Leuven, Belgium}

\affiliation[b]{Université Paris-Saclay, CNRS, CEA, \\
	Institut de physique théorique, 91191, Gif-sur-Yvette, France}

\emailAdd{nikolay.bobev@kuleuven.be}
\emailAdd{junho.hong@kuleuven.be}
\emailAdd{valentin.reys@ipht.fr}

\abstract{We use holography in conjunction with recent results from supersymmetric localization to compute certain thermal observables for 3d $\mathcal{N}=2$ holographic SCFTs arising on the worldvolume of $N$ M2-branes. We obtain results for the thermal free energy density on $S^1 \times \mathbb{R}^2$, the Casimir energy on $T^{2} \times \mathbb{R}$, and the three leading coefficients in the large temperature limit of the free energy on $S^1\times S^2$ valid to subleading order in the large $N$ limit. As a byproduct of our holographic analysis we also present a conjecture for the structure of the large temperature expansion of the thermal free energy of general 3d CFTs on $S^1\times S^2$.}

\begin{document}
	
\maketitle

\section{Introduction}
\label{sec:intro}

Thermal observables in strongly interacting QFTs are of great interest for a variety of physical applications, ranging from condensed matter physics to string theory, but are notoriously difficult to compute with conventional methods. The structure of thermal correlation functions in theories enjoying conformal invariance in their vacuum state is more constrained and is therefore a natural starting point to better understand thermal physics in QFTs. Recently a number of different techniques have been employed to gain insight into thermal correlation functions in CFTs including ideas from the conformal bootstrap, holography, as well as certain limits of thermal partition functions on compact Euclidean manifolds, see \cite{El-Showk:2011yvt,Katz:2014rla,Iliesiu:2018fao,Rodriguez-Gomez:2021pfh,Rodriguez-Gomez:2021mkk,Luo:2022tqy,Benjamin:2023qsc,Marchetto:2023fcw,Diatlyk:2023msc} for a selection of recent works relevant to the present context.

Our goal here is to build on these recent developments and leverage new results from supersymmetric localization, holography, and higher-derivative supergravity to compute several thermal observables for holographic SCFTs. Our main focus will be on 3d $\mathcal{N}=2$ SCFTs arising on the worldvolume of $N$ coincident M2-branes. The prototypical example of this class of theories is the ABJM theory  \cite{Aharony:2008ug}. We will study the so-called M-theory limit of these SCFTs which is obtained by keeping the Chern-Simons level of the gauge group(s) fixed and taking $N$ to infinity. In this limit the theory is holographically dual to 11d supergravity on a background that is asymptotic to AdS$_4\times {\rm SE}_7$, where ${\rm SE}_7$ is an appropriate 7d Sasaki-Einstein manifold, the Lens space $S^7/\mathbb{Z}_k$ being the most studied example corresponding to the ABJM theory. Thermal correlation functions of local operators in this class of theories admit a large $N$ expansion where the leading term is of order $N^{3/2}$ and the first subleading term is generally expected to be of order $N^{1/2}$. Computing these leading and subleading terms in the large $N$ expansion of thermal correlators will be the main goal of this work.

One of the observables we focus on is the thermal free energy density, $f_{\mathcal{T}}$, of the CFT placed on $S^1\times \mathbb{R}^2$ with thermal boundary conditions on the $S^1$ of circumference $\beta$ determined by the inverse temperature. As we review in Section~\ref{sec:thermal}, $f_{\mathcal{T}}$ is proportional to the coefficient of the one-point function of the energy-momentum tensor of the CFT at finite temperature. Moreover, the same quantity $f_{\mathcal{T}}$ also determines the leading behavior of the Casimir energy of the CFT when placed on $T^2 \times \mathbb{R}$ in the limit where the modular parameter of the torus takes large imaginary values. It is straightforward to calculate $f_{\mathcal{T}}$ for free theories but much more challenging for strongly coupled CFTs. For CFTs that admit a large $N$ limit, like the ${\rm O}(N)$ vector model, one can calculate $f_{\mathcal{T}}$ to leading and subleading orders in the large $N$ expansion, see~\cite{Chubukov:1993aau,Sachdev:1993pr,Diatlyk:2023msc}. Another class of large $N$ theories for which $f_{\mathcal{T}}$ can be calculated are the ones that admit an explicit holographic dual description in string or M-theory, see \cite{Gubser:1998nz,Kovtun:2008kw}. In these models one can determine $f_{\mathcal{T}}$ by calculating the regularized on-shell action of the AdS soliton solution of 4d gravity which is the relevant background dual to a thermal 3d CFT on $S^1\times \mathbb{R}^2$  \cite{Horowitz:1998ha}. This result is well-known but is valid only to leading order in the supergravity approximation, i.e. leading order in the large $N$ limit. Going beyond this leading order is challenging since one has to include higher-derivative corrections to supergravity. One approach is to adopt a bottom-up perspective and study the effects of general higher-derivative corrections to holographic thermal observables, see for instance \cite{Buchel:2009sk}. This has the disadvantage that one lacks detailed information about the relation between the higher-derivative gravitational couplings and the microscopic parameters in the dual CFTs. As shown in \cite{Bobev:2020egg,Bobev:2021oku} one can also adopt a top-down perspective, determine the four-derivative action of 4d $\mathcal{N}=2$ minimal gauged supergravity and explicitly calculate the gravitational couplings in this theory in terms of the parameters in the dual 3d $\mathcal{N}=2$ SCFT. This can be done for a large class of SCFTs in the ABJM ``universality class'' by judiciously leveraging results from supersymmetric localization. The basic idea is to calculate supersymmetric partition functions of the 3d $\mathcal{N}=2$ SCFT of interest on several different compact Euclidean manifolds. These results should then be equal to the regularized on-shell action of an appropriate supersymmetric Euclidean solution in the four-derivative supergravity theory. Comparing the explicit supersymmetric localization results with the gravitational on-shell actions then allows one to determine explicitly the 
gravitational couplings in the supergravity theory in terms of the CFT parameters. While supersymmetry is essential in applying this procedure, once one has obtained the 4d higher-derivative gravitational action with its couplings fixed in terms of the dual microscopic data, one can use it to compute non-supersymmetric thermal observables. This is the essence of the method we employ in this work in order to compute $f_{\mathcal{T}}$ to order $N^{1/2}$ for a class of 3d $\mathcal{N}=2$ SCFTs.

Using the same approach we also study the higher-derivative on-shell action of the 4d AdS-Kerr solution. The relevant thermal observable is the free energy of the dual 3d SCFT on $S^1 \times S^2$ in the presence of a non-trivial chemical potential for the angular momentum on $S^2$. The large temperature limit of this free energy is of interest since, among other things, it encodes the asymptotic growth of states in the 3d CFT in radial quantization, akin to the Cardy behavior of 2d CFTs, see~\cite{Benjamin:2023qsc} as well as \cite{Bhattacharyya:2007vs,Banerjee:2012iz,Jensen:2012jh,DiPietro:2014bca,Shaghoulian:2015lcn} for related work. The leading term in the large temperature expansion is proportional to the quantity $f_{\mathcal{T}}$ introduced above. As discussed in~\cite{Benjamin:2023qsc}, the subleading term is controlled by two constants $c_{1,2}$ which contain dynamical information about the CFT at hand. Using the approach outlined above, we compute $c_{1,2}$ in terms of the M2-brane SCFT microscopic data up to order $N^{1/2}$. Moreover, we use the result for the on-shell action of the 4d AdS-Kerr solution together with the free energy of free 3d scalars and fermions to arrive at a conjecture for the general structure of the high temperature limit of the $S^1 \times S^2$ thermal free energy for any 3d CFT and to all orders in the temperature.

We continue our presentation in the next section with a short overview of the CFT thermal observables that we study in this work. We proceed in Section~\ref{sec:AdSsolHD} with a discussion on how to compute these observables holographically using recent results from four-derivative gauged supergravity. In Section~\ref{sec:M2branes} we apply these holographic results in conjunction with supersymmetric localization to 3d SCFTs and derive explicit expressions for the thermal observables at leading and subleading order in the large $N$ expansion. In Section~\ref{sec:Sthermal} we combine the holographic analysis of the AdS-Kerr on-shell action with explicit results for 3d free scalars and fermions to propose a conjecture for the general form of the thermal free energy of 3d CFTs on $S^1\times S^2$. We conclude in Section~\ref{sec:discussion} with a short discussion.

\section{Thermal CFT observables}
\label{sec:thermal}

We begin this section with a quick review of various CFT observables that will be of interest in this paper, together with their interrelations. When a $d$-dimensional CFT is placed on $S^1_{\beta}\times \mathbb{R}^{d-1}$, with $\beta=1/T$ the circumference of the thermal circle with temperature $T$, the one-point functions of primary operators no longer vanish due to the broken conformal invariance, see~\cite{Iliesiu:2018fao} for a summary. Nevertheless, their form is significantly constrained. For example, for a scalar operator $\mathcal{O}$ of conformal dimension $\Delta_{\mathcal{O}}$, this thermal one-point function is given by
\begin{equation}
\langle\mathcal{O}\rangle_{\beta} = \frac{b_{\mathcal{O}}}{\beta^{\Delta_{\mathcal{O}}}}\,,
\end{equation}
for some $\beta$-independent coefficient $b_{\mathcal{O}}$. In addition, the one-point functions of all conformal descendants have to vanish due to translation invariance on $\mathbb{R}^{d-1}$. The coefficients $b_{\mathcal{O}}$ are not determined by conformal invariance and encode dynamical information about the CFT. In the thermal context, they should therefore be treated as a supplement to the usual local CFT data $\{\Delta_i,C_{ijk}\}$ of conformal dimensions and OPE coefficients that determine the correlation functions on $\mathbb{R}^{d}$.

A special role in every CFT is played by the energy momentum operator $\mathcal{T}_{\mu\nu}$. In particular, the energy density of the CFT at finite temperature is given by
\begin{equation}
\label{eq:bT-def}
E(\beta) = - \langle \mathcal{T}^{00}\rangle_{\beta} = - \frac{d-1}{d}\,\frac{b_{\mathcal{T}}}{\beta^{d}}\,.
\end{equation}
Clearly by positivity of energy one finds that in these conventions unitary CFTs have $b_{\mathcal{T}} <0$.\footnote{If the theory is topological, the coefficient $b_\mathcal{T}$ may in fact vanish.} On the other hand, the free energy density of the theory is fixed by dimensional analysis to take the form
\begin{equation}
\label{eq:F-fT}
F_{S^1 \times \mathbb{R}^{d-1}} = \frac{f_{\mathcal{T}}}{\beta^{d}}\, ,
\end{equation}
for some dimensionless coefficient $f_\mathcal{T}$. Using the thermodynamic relations
\begin{equation}
F = E- T S\,, \qquad S = - \frac{dF}{dT } \,,
\end{equation}
one finds that $f_\mathcal{T}$ is related to the one-point function coefficient $b_\mathcal{T}$ as
\begin{equation}
\label{eq:f-b}
f_{\mathcal{T}} = \frac{b_{\mathcal{T}}}{d}\,.
\end{equation}

The quantity $b_{\mathcal{T}}$ is hard to compute for most interacting CFTs, although some known results are summarized in \cite{Iliesiu:2018fao}. As an example, the free energy of a single free scalar field can be computed directly to obtain
\begin{equation}
\label{eq:bT-free}
b_{\mathcal{T}}^{s} = - \frac{d\,\zeta(d) \Gamma(d/2)}{\pi^{d/2}}\, .
\end{equation}
In the planar limit of the critical 3d $O(N)$ model one finds that to leading order in $N$~\cite{Chubukov:1993aau,Sachdev:1993pr},
\begin{equation}
b_{\mathcal{T}}^{O(N)} \approx \frac{4N}{5} b_{\mathcal{T}}^{s} \approx -0.45915 N\,,
\end{equation}
where we have used that in three dimensions, $b_{\mathcal{T}}^{s} = - \frac{3\zeta(3)}{2\pi} \approx -0.57394$.\\

Another observable of interest is the Casimir energy of a CFT placed on $T^2\times \mathbb{R}^{d-2}$ which was recently discussed in~\cite{Luo:2022tqy}. By dimensional analysis, this quantity must take the form $E_{\text{cas}} = -\mathcal{E}(\tau) \mathcal{A}^{(2-d)/2}$ where $\mathcal{A}$ is the torus area and $\tau$ the modular parameter on $T^2$. By exploiting modular covariance, it can be shown that
\begin{equation}
\label{eq:g1-g2}
\lim_{\tau\to\mathrm{i}\infty}\mathcal{E}(\tau) = g_1\,\text{Im}(\tau)^{d/2} + g_2\,\text{Im}(\tau)^{1-d/2} + \ldots
\end{equation}
where the dots denote exponentially suppressed terms as $\text{Im}(\tau)$ becomes large. This limit corresponds to taking one of the compact directions to be large, and as a consequence the coefficient controlling the leading term in \eqref{eq:g1-g2} can be related to the one-point function coefficient $b_\mathcal{T}$ defined previously as follows~\cite{Luo:2022tqy}
\begin{equation}
\label{eq:thermal-rel}
g_1 =  - f_{\mathcal{T}} = - \frac{b_{\mathcal{T}}}{d}\,.
\end{equation}
The second coefficient $g_2 \geq 0$ gives a measure of the gapless degrees of freedom in the effective field theory (EFT) obtained by KK reduction of the CFT on the small cycle inside the $T^2$~\cite{Luo:2022tqy}. Consequently, one has $g_2 = 0$ when the EFT is gapped.\\

Computing the above observables for a given strongly interacting CFT is a non-trivial task. When the theory is holographic however, one can leverage the AdS/CFT correspondence to obtain non-trivial results from the dual bulk gravitational theory. Our goal in the next section will be to study the one-point function coefficient $b_\mathcal{T}$ (or equivalently $g_1$ or $f_{\mathcal{T}}$ using \eqref{eq:thermal-rel}) and the Casimir coefficient $g_2$ for holographic SCFTs arising from M-theory beyond the leading supergravity approximation. For concreteness, we will be primarily interested in three-dimensional CFTs and thus we will often set $d=3$ in the expressions above.

\section{AdS backgrounds and HD supergravity}
\label{sec:AdSsolHD}

The free energy of a holographic CFT on $S^1_{\beta}\times \mathbb{R}^{d-1}$ or its Casimir energy density defined in \eqref{eq:g1-g2} can be computed by using the standard holographic dictionary and evaluating the on-shell action of the relevant asymptotically AdS backgrounds. Using the recent four-dimensional higher-derivative (HD) supergravity results in~\cite{Bobev:2020egg,Bobev:2021oku} then gives access to holographic thermal quantities beyond the leading supergravity approximation, as we show below.\\ 

Let us first briefly summarize the HD results. The dynamics of the stress-tensor of 3d $\mathcal{N}=2$ SCFTs to leading order in the large $N$ limit is captured holographically by minimal supergravity, whose two-derivative bosonic action is simply that of the Einstein-Maxwell theory. In our discussion we will be interested in working at finite temperature and vanishing charge density so the $\U(1)$ gauge field will not play a role and we will no longer discuss it explicitly. The two-derivative action of interest in Euclidean signature thus reads 
\begin{equation}
S_{2\partial} = -\frac{1}{16\pi G_N}\int d^4x \sqrt{g}\,\Bigl[R + \frac{6}{L^2}\Bigr] \, .
\end{equation}
In addition, one can consider four-derivative couplings made up of contractions of the curvature tensors. Such couplings are most easily constructed in the superconformal framework, see \cite{Lauria:2020rhc} for a review. Upon eliminating the auxiliary fields, the resulting Euclidean bosonic HD action (with vanishing Maxwell field) is given by~\cite{Bobev:2020egg,Bobev:2021oku}
\begin{equation}
\label{eq:4der-action}
S_{4\partial} = S_{2\partial} + (\lambda_1 - \lambda_2)\int d^4x\sqrt{g}\,(C_{\mu\nu\rho\sigma})^2 + \lambda_2 \int d^4x\sqrt{g}\,\Bigl[(R_{\mu\nu\rho\sigma})^2 - 4(R_{\mu\nu})^2 + R^2\Bigr] \, ,
\end{equation}
where $C_{\mu\nu\rho\sigma}$ is the Weyl tensor and $\lambda_{1,2}$ are constant dimensionless coefficients.\footnote{The four-derivative coefficients were denoted by $c_{1,2}$ in \cite{Bobev:2020egg,Bobev:2021oku}. We denote them with $\lambda_{1,2}$ here to avoid confusion with the coefficients in the thermal effective action in \eqref{eq:S-th} below.} One crucial property of the HD action \eqref{eq:4der-action} is that any solution to the two-derivative Einstein equations also solves the equations of motion of the four-derivative theory, for any value of the coefficients $\lambda_{1,2}$. We stress that this property holds regardless of the amount of supersymmetry preserved by the two-derivative solution. This allows us to consider HD corrections to non-supersymmetric minimal supergravity backgrounds, and in particular to the backgrounds dual to the observables reviewed in Section~\ref{sec:thermal}. Furthermore, as shown in~\cite{Bobev:2020egg,Bobev:2021oku}, there exists a unique set of HD counter-terms that ensure that the variational problem based on \eqref{eq:4der-action} is well-defined, and which provide the necessary and sufficient divergent terms to regulate the UV divergences when evaluating the HD action on-shell. The result of this holographic renormalization procedure is given by the following four-derivative on-shell action
\begin{equation}
\label{eq:4der-os}
I_{4\partial} = \Bigl[1 + \frac{64\pi G_N}{L^2}(\lambda_2 - \lambda_1)\Bigr]\,I_{2\partial} + 32\pi^2\,\lambda_1\,\chi \, ,
\end{equation}
where $L$ is the length scale of the asymptotically AdS solution, $I_{2\partial}$ is the two-derivative renormalized on-shell action, and $\chi$ is the regularized Euler characteristic of the 4d Euclidean asymptotically AdS space. We now proceed to apply this general result to extract holographic thermal observables.

\subsection{The AdS soliton dual to $S^1_\beta \times \mathbb{R}^2$}
\label{subsec:AdSSoliton}

In general dimension, the metric of the Lorentzian AdS soliton is obtained by a double analytic continuation of the near-extremal $p$-brane metric, see~\cite{Horowitz:1998ha}. Asymptotically, the topology is $S_\beta^1 \times \mathbb{R}^{1,p-1}$. Since we are interested in thermal observables of three-dimensional CFTs, we set $p=2$ in what follows. To obtain a Euclidean background one needs to Wick rotate the time direction in $\mathbb{R}^{1,1}$ and the resulting metric reads 
\begin{equation}
\label{eq:soliton-S1}
ds^2 = \frac{r^2}{L^2}\Bigl[\Bigl(1 - \frac{r_0^3}{r^3}\Bigr)dt_E^2 + dx^2 + dy^2\Bigr] + \frac{L^2}{r^2}\Bigl(1 - \frac{r_0^3}{r^3}\Bigr)^{-1}dr^2 \, .
\end{equation}
where $t_E$ parametrizes the asymptotic $S_\beta^1$ circle, $(x,y)$ are flat coordinates on $\mathbb{R}^2$ and $r\geq r_0$. The boundary is located at $r \to \infty$. To avoid a conical singularity at $r=r_0$ one must identify $r_0 = 4\pi L^2/3\beta$ where $\beta$ is the circumference of the $S_\beta^1$ circle. Using the standard holographic counterterms one can evaluate the two-derivative on-shell action to find
\begin{equation}
I^{S^1 \times \mathbb{R}^2}_{2\partial} = -\frac{r_0^3\,\text{vol}(\mathbb{R}^2)\beta}{16\pi G_N L^4} \, .
\end{equation}
Furthermore, it is straightforward to check that the regularized Euler characteristic of the space vanishes. Using \eqref{eq:4der-os}, we therefore obtain the following corrected on-shell action at the four-derivative level:
\begin{equation}
I^{S^1 \times \mathbb{R}^2}_{4\partial} = -\Bigl[1 + \frac{64\pi G_N}{L^2}(\lambda_2 - \lambda_1)\Bigr]\frac{r_0^3\,\text{vol}(\mathbb{R}^2)\beta}{16\pi G_N L^4} \, .
\end{equation}
From this, we deduce that the HD-corrected free energy density of the AdS soliton is given by
\begin{equation}
\label{eq:E-soliton}
F_{S^1 \times \mathbb{R}^2} = \frac{1}{\text{vol}(\mathbb{R}^2)\beta}\,I^{S^1 \times \mathbb{R}^2}_{4\partial} = -\frac{4\pi^2 L^2}{27G_N\beta^3} + \frac{256\pi^3}{27\beta^3}(\lambda_1 - \lambda_2) \, .
\end{equation}
According to \eqref{eq:F-fT}, this leads to the HD-corrected thermal observable
\begin{equation}
\label{eq:fT-soliton}
f_{\mathcal{T}} = -\frac{4\pi^2 L^2}{27G_N} + \frac{256\pi^3}{27}(\lambda_1 - \lambda_2) \, .
\end{equation}
In the next section, we will use the holographic dictionary to write this quantity in terms of CFT data for some specific examples of 3d $\mathcal{N}=2$ SCFTs. Before doing so, we discuss other minimal supergravity backgrounds dual to other CFT observables.

\subsection{The AdS soliton dual to $T^2 \times \mathbb{R}$}

The AdS soliton dual to the CFT background $T^2 \times \mathbb{R}$ amounts to a different choice of compact cycles compared to \eqref{eq:soliton-S1}. The Lorentzian metric now reads~\cite{Horowitz:1998ha,Luo:2022tqy}
\begin{equation}
\label{eq:soliton-T2}
ds^2 = -\frac{r^2}{L^2}dt^2 + \frac{L^2}{r^2}\Bigl(1 - \frac{r_0^3}{r^3}\Bigr)^{-1}dr^2 + \frac{r^2 \beta^2}{L^2}\Bigl[\Bigl(1 - \frac{r_0^3}{r^3}\Bigr)(dx_1 + \tau_1dx_2)^2 + \tau_2^2 dx^2_2\,\Bigr] \, ,
\end{equation}
where we parametrize the $T^2 $ with coordinates $x_1 \sim x_1 + 1$ and $x_2 \sim x_2 + 1$ and use $\tau = \tau_1 + \mathrm{i}\tau_2$ to denote its modular parameter. The radial coordinate $r$ is as in the previous subsection. The renormalized two-derivative on-shell action of this soliton is given by
\begin{equation}
I_{2\partial}^{T^2 \times \mathbb{R}} = -\frac{r_0^3\,\text{vol}(\mathbb{R})}{16\pi G_N L^4}\,\mathcal{A} \, ,
\end{equation}
where $\mathcal{A} = \beta^2\tau_2$ is the torus area. The Euler characteristic of the space \eqref{eq:soliton-T2} remains zero, and we find that the energy density of this AdS soliton is given by
\begin{equation}
\frac{1}{\text{vol}(\mathbb{R})}\,I_{4\partial}^{T^2 \times \mathbb{R}} = -\Bigl[\frac{4\pi^2 L^2}{27 G_N \beta^3} - \frac{256\pi^3}{27\beta^3} (\lambda_1 - \lambda_2)\Bigr]\mathcal{A} \, ,
\end{equation}
which, upon using the holographic relation $E_{\rm cas} = I_{4\partial}^{T^2 \times \mathbb{R}} / \text{vol}(\mathbb{R})$, leads to the HD-corrected Casimir energy density
\begin{equation}
\mathcal{E}(\tau) = \Bigl[\frac{4\pi^2 L^2}{27 G_N} - \frac{256\pi^3}{27}(\lambda_1 - \lambda_2)\Bigr]\,\tau_2^{3/2} \, .
\end{equation}
Comparing to \eqref{eq:g1-g2}, we thus find
\begin{equation}\label{eq:g1-g2holo4der}
g_1 = -f_\mathcal{T} \, , \qquad g_2 = 0 \, ,
\end{equation}
where $f_\mathcal{T}$ is given in \eqref{eq:fT-soliton}. The first relation is expected from \eqref{eq:thermal-rel}. We also find that $g_2$ vanishes even after including HD corrections, which suggests that the EFT obtained after KK reduction of the CFT on the small thermal cycle inside $T^2$ remains gapped after incorporating the first $1/N$ correction in the planar limit. Lastly, no term independent of the torus complex structure is generated since the Euler characteristic of the soliton \eqref{eq:soliton-T2} vanishes. This is compatible with the general results about the $T^2$ Casimir energy in \cite{Luo:2022tqy}.

\subsection{The AdS-Kerr background}
\label{sec:AdS-Kerr}

We now consider another supergravity background that is dual to 3d CFTs placed on $S^1_\beta \times S^2$. The grand canonical partition function on such a space captures the spectrum of the CFT, and can be elegantly computed using the formalism of the so-called thermal effective action, see~\cite{Benjamin:2023qsc} as well as \cite{Bhattacharyya:2007vs,Banerjee:2012iz,Jensen:2012jh,DiPietro:2014bca,Shaghoulian:2015lcn} for related work. In 3d, we have 
\begin{equation}
\label{eq:Sth-def}
Z_{S^1 \times S^2}(T,\Omega) = \text{Tr}\Bigl[e^{-\beta D + \mathrm{i}\beta\,\Omega\,M}\Bigr] \, ,
\end{equation}
where $D$ is the dilatation operator, $M$ is the generator of the Cartan subalgebra of the rotation group $\SO(3)$, and $\Omega$ is the spin fugacity. In the large temperature, or small $\beta$, limit\footnote{We fix the radius of the $S^2$ to be $1$ which sets the scale in the problem. The high temperature limit therefore corresponds to $\beta \ll 1$.} the trace can be approximated by a saddle-point as
\begin{equation}\label{eq:Sthsaddle}
F_{S^1 \times S^2}(T,\Omega) =-\log Z_{S^1 \times S^2}(T,\Omega) \approx S_{\text{th}}(T,\Omega) \, .
\end{equation}
where on the right-hand side we have used the on-shell value of the 4d thermal effective action $S_{\text{th}}$ that encodes thermal and spin information about the CFT. Using $S_{\text{th}}$ it can be shown that the thermal free energy on $S^1 \times S^2$ admits the following large temperature expansion~\cite{Benjamin:2023qsc}:
\begin{equation}
\label{eq:S-th}
F_{S^1 \times S^2}(T,\Omega) = \frac{4\pi}{1+\Omega^2}\Bigl[f_{\mathcal{T}}\,T^2 + 2\,c_1\,(1 + \Omega^2) + \frac83\,c_2\,\Omega^2 + \mathcal{O}(T^{-2})\Bigr] \, .
\end{equation}
Here, the leading term is determined by $f_{\mathcal{T}}$ defined in~\eqref{eq:F-fT}. The first subleading term is controlled by two coefficients $c_1$ and $c_2$ which, in the holographic context, can be determined from a supergravity computation as follows. The relevant gravitational dual background is the Euclidean AdS-Kerr black hole solution:
\begin{equation}
\label{eq:Kerr-metric}
ds^2 = \frac{\Delta_r}{W}\Bigl(dt_E + \frac{\alpha}{\Xi}\sin^2\theta\,d\phi\Bigr)^2 + W\Bigl(\frac{dr^2}{\Delta_r} + \frac{d\theta^2}{\Delta_\theta}\Bigr) + \frac{\Delta_\theta\sin^2\theta}{W}\Bigl(\alpha\,dt_E - \frac{r^2 - \alpha^2}{\Xi}d\phi\Bigr)^2 \, ,
\end{equation}
where
\begin{equation}
\begin{split}
W(r,\theta) =&\; r^2 - \alpha^2\cos^2\theta \, , \qquad \Xi = 1 + \frac{\alpha^2}{L^2} \, , \\
\Delta_r(r) =&\; (r^2 - \alpha^2)\Bigl(1 + \frac{r^2}{L^2}\Bigr) - 2mr \, , \qquad \Delta_\theta(\theta) = 1 + \frac{\alpha^2}{L^2}\cos^2\theta \, .
\end{split}
\end{equation}
The space has an outer horizon located at the largest real root of $\Delta_r(r_+) = 0$. To ensure regularity of the Euclidean time circle at $r=r_+$, we must identify $t_E \sim t_E + \beta$ where
\begin{equation}
\label{eq:beta-Kerr}
\beta = \frac{2\pi L^2(r_+^2 - \alpha^2)}{2r_+^3 + r_+(L^2 - \alpha^2) - mL^2} \, .
\end{equation}
The angular velocity should be measured relative to a non-rotating frame at infinity and reads
\begin{equation}
\label{eq:omega-Kerr}
\omega = \alpha\,\frac{1+r_+^2L^{-2}}{r_+^2 - \alpha^2} \, .
\end{equation}
The black hole parameters $(m,\alpha)$ that characterize the solution are related to its mass and angular momentum. The two-derivative regularized on-shell action is given by~\cite{Cassani:2019mms,Bobev:2021oku}\footnote{We have corrected a typo in Eq. (3.42) of~\cite{Bobev:2021oku}.}
\begin{equation}
\label{eq:I-Kerr-2der}
I^{S^1\times S^2}_{2\partial} = \frac{\beta}{2G_N\Xi}\Bigl(m - \frac{r_+}{L^2}(r_+^2-\alpha^2)\Bigr) \, .
\end{equation}
We can use the equation $\Delta_r(r_+)=0$ to eliminate $m$ in \eqref{eq:beta-Kerr}, use \eqref{eq:omega-Kerr} to solve for $\alpha$ as a function of $r_+$ and $\omega$, and use the resulting equation to solve for $r_+$ in the large temperature limit to find the expansion 
\begin{equation}
	\label{eq:r+-beta:Omega}
	r_+=\fft{4\pi L^2}{3\beta}-\fft{1+2L^2\omega^2}{4\pi}\beta-\fft{3(1-2L^2\omega^2-2L^4\omega^4)}{64\pi^3L^2}\beta^3+\mathcal{O}(\beta^5)\,.
\end{equation}
Using this together with the solution for $\alpha$ as a function of $r_+$ and $\omega$ from \eqref{eq:omega-Kerr} in~\eqref{eq:I-Kerr-2der} leads to the following high-temperature behavior of the on-shell action:\footnote{We note that we keep $\omega$ fixed while taking the small $\beta$ limit. This is in harmony with the high temperature limit in the dual CFT which is taken while keeping $\Omega$ fixed.}
\begin{equation}\label{eq:I2derKerrhighT}
I^{S^1\times S^2}_{2\partial} = -\frac{16\pi^3 L^2}{27 G_N}\frac{(\beta/L)^{-2}}{1 + L^2\omega^2}\Bigl[1 - \frac{9\,(2 + L^2\omega^2)}{16\pi^2}\,(\beta/L)^2 + \mathcal{O}\bigl((\beta/L)^4\bigr)\Bigr] \, .
\end{equation}
It can further be shown that the regularized Euler characteristic of \eqref{eq:Kerr-metric} is $\chi = 2$, see~\cite{Bobev:2021oku}. Comparing the resulting four-derivative regularized on-shell action \eqref{eq:4der-os} with the free energy in \eqref{eq:S-th}, we arrive at the following identification:\footnote{We note that \eqref{eq:S-th} is written in terms of dimensionless temperature and spin fugacity, whereas the gravitational on-shell action is naturally expressed in terms of dimensionful quantities. This explains the factors of $L$ in the combinations $\beta/L$ and $L\,\omega$ appearing in \eqref{eq:I2derKerrhighT}. To translate between the two conventions we can simply set $L=1$ and $\omega=\Omega$.}
\begin{equation}
\label{eq:thermal-grav}
\begin{split}
f_\mathcal{T} =&\; -\frac{4\pi^2L^2}{27G_N} + \frac{256\pi^3}{27}(\lambda_1 - \lambda_2) \, , \\
c_1 =&\; \frac{L^2}{12G_N} + \frac{8\pi}{3}(\lambda_1 + 2\lambda_2) \, , \\
c_2 =&\; -\frac{L^2}{32G_N} + 2\pi(\lambda_1 - \lambda_2) \, .
\end{split}
\end{equation}
We note that the result for $f_\mathcal{T}$ is the same as the one derived from the AdS soliton \eqref{eq:fT-soliton}, which is a nice consistency check. The HD-corrected effective action coefficients $c_{1,2}$ are new. Since they are obtained from a bulk computation these thermal quantities are written in terms of gravitational variables, and we will now make use of the holographic dictionary to write them in terms of CFT data for some specific AdS/CFT examples.

\section{M2-brane holography}
\label{sec:M2branes}

In this section we explain how holography can be used to write the thermal quantities \eqref{eq:thermal-grav} in a CFT language. We consider 3d $\mathcal{N}\geq 2$ SCFTs arising from the low-energy limit of a stack of $N$ M2-branes probing a conical singularity. In this case, we expect the gravitational parameters to scale as
\begin{equation}
\label{eq:grav-N}
\frac{L^2}{2G_N} = A\,N^{3/2} + a\,N^{1/2} \, , \qquad 32\pi \lambda_i = v_i\,N^{1/2} \, ,
\end{equation}
in the large $N$ limit~\cite{Camanho:2014apa,Bobev:2021oku}. The $N$-independent quantities $(A,a,v_i)$ can be determined as follows. First, according to \eqref{eq:4der-os}, the four-derivative regularized on-shell action of any minimal supergravity background $\mathbb{S}$ takes the form
\begin{equation}
\label{eq:4der-os-N}
I_{4\partial} = \pi\mathcal{F}(\mathbb{S})\Bigl(A\,N^{3/2} + (a + v_2)\,N^{1/2}\Bigr) - \pi\bigr(\mathcal{F}(\mathbb{S}) - \chi(\mathbb{S})\bigl) \,v_1\,N^{1/2} + \mathcal{O}(\log N) \, ,
\end{equation}
where we defined
\begin{equation}
I_{2\partial} = \frac{\pi L^2}{2G_N}\,\mathcal{F}(\mathbb{S}) \, .
\end{equation}
As shown in~\cite{Bobev:2020egg,Bobev:2021oku}, we can then use results for the first two terms in the large $N$ limit of supersymmetric observables in the dual SCFTs to obtain the values of the combinations $a + v_2$ and $v_1$. One such result is available for the topologically twisted index of $\mathcal{N}\geq 2$ SCFTs, which is the supersymmetric index obtained by putting the SCFT on $S^1\times \Sigma_{\mathfrak{g}}$ while turning on a background flux for the exact R-symmetry in order to preserve supersymmetry via a (partial) twist along the Riemann surface $\Sigma_\mathfrak{g}$~\cite{Benini:2015noa,Benini:2016hjo}. The path integral representation of this index can be computed exactly using localization, and the resulting matrix model can be studied very precisely using the Bethe Ansatz formulation and numerical techniques, as shown in~\cite{Bobev:2022jte,Bobev:2022eus,Bobev:2023lkx}. The upshot is that its large $N$ limit reads
\begin{equation}
\label{eq:TTI}
\frac{\log Z_{S^1 \times \Sigma_\mathfrak{g}}}{\mathfrak{g} - 1} = \pi\,\alpha\,\Bigl((N - B)^{3/2} + C\,(N - B)^{1/2}\Bigr) + \frac12\log(N - B) + \mathcal{O}(e^{-\sqrt{N}}) \, ,
\end{equation}
where the $N$-independent coefficients $(\alpha,B,C)$ depend on the SCFT of interest, and we will give some examples shortly. The gravity background dual to the topologically twisted index is a Euclidean black saddle known as the Romans solution~\cite{Romans:1991nq,Cacciatori:2009iz,Gauntlett:2001qs,Bobev:2020pjk,BenettiGenolini:2019jdz}, for which $\mathcal{F}=1-\mathfrak{g}$ and $\chi = 2-2\mathfrak{g}$. Comparing \eqref{eq:4der-os-N} for this background with \eqref{eq:TTI}, we obtain
\begin{equation}
A = \alpha \, , \qquad a + v_1 + v_2 = \alpha\Bigl(C - \frac32\,B\Bigr) \, .
\end{equation}
Note that the twisted index alone is not sufficient to disentangle the contributions to the $\mathcal{O}(N^{1/2})$ coefficients entering the holographic dictionary \eqref{eq:grav-N}. For this, we will use another SCFT observable known as the superconformal index. This quantity is obtained by considering the supersymmetric index of the SCFT when placed on $S^1 \times_\omega S^2$. Here, $\omega$ denotes the $S^2$ angular momentum fugacity. The superconformal index can be studied in the Cardy-like limit of small $\omega$, which physically corresponds to the limit where the size of the $S^1$ circle is much smaller than the radius of the $S^2$.\footnote{Note that the superconformal index on $S^1\times S^2$ is a different physical quantity from the thermal partition function on $S^1\times S^2$ discussed in Section~\ref{sec:AdS-Kerr} and Section~\ref{sec:Sthermal} due to the supersymmetric periodicity conditions imposed on the $S^1$.} In this regime, it was shown in~\cite{Bobev:2022wem} that the index takes the form
\begin{align}
\label{eq:SCI}
-\log Z_{S^1\times_\omega S^2} =&\; \frac{\pi\alpha}{2\omega}\,(N - B)^{3/2} \\
&\;+ \pi\alpha\,\Bigl((N - B)^{3/2} + C(N - B)^{1/2}\Bigr) + \frac12\log(N - B) + \mathcal{O}(e^{-\sqrt{N}},\omega) \, , \nonumber
\end{align}
where the coefficients $(\alpha,B,C)$ are the same as the ones appearing in the topologically twisted index \eqref{eq:TTI}. We note that in~\cite{Bobev:2022wem} this relation between the two different indices was shown analytically for the $\mathcal{N}=6$ ABJM theory and the 3d $\mathcal{N}=4$ SYM theory known as ADHM (or the $N_f$ model). We have recently improved on this analysis and have shown that this relation is valid for general 3d $\mathcal{N}=2$ SCFTs \cite{progress} and we can therefore employ it safely in our current discussion.

The gravity dual to the superconformal index is given by the supersymmetric limit of the Euclidean Kerr-Newman-AdS solution, for which $\mathcal{F} = (\omega+1)^2/(2\omega)$ and $\chi = 2$. Using this and comparing \eqref{eq:4der-os-N} with \eqref{eq:SCI} yields $A = \alpha$ and $a - v_1 + v_2 = -3\alpha B/2$. Combining this with the data obtained from the topologically twisted index, we arrive at the holographic dictionary
\begin{equation}
\label{eq:holo-dict}
A = \alpha \qquad a + v_2 = \frac{\alpha}{2}\,(C - 3B) \, , \qquad v_1 = \frac{\alpha}{2}\,C \, .
\end{equation}

The coefficients $(\alpha,B,C)$ in \eqref{eq:holo-dict} have been obtained from localization in a variety of $\mathcal{N}\geq 2$ SCFTs, see~\cite{Bobev:2022jte,Bobev:2022eus,Bobev:2023lkx}. These theories describe the world-volume of M2-branes placed at the tip of a cone over various seven-dimensional Sasaki-Einstein manifolds $Y_7$ or orbifolds thereof. We have collected this data in Table~\ref{tab:loc}.\footnote{Another holographic CFT for which we can compute these quantities is the so-called mABJM theory which is obtained from the ABJM model by adding a superpotential mass term \cite{Jafferis:2011zi,Bobev:2018uxk,Bobev:2018wbt}. The AdS$_4$ 11d supergravity dual to this CFT is not of the Freund-Rubin type and was constructed in \cite{Corrado:2001nv}. For this model we find the following values: $\alpha = \frac{4 \sqrt{2 k}}{9\sqrt{3}}$, $B= \frac{k}{24} - \frac{5}{6 k}$, $C= -\frac{9}{2k}$, see ~\cite{Bobev:2022jte,Bobev:2022eus,Bobev:2023lkx}. This can be used in \eqref{eq:bTc1c2M2} to obtain the thermal observables $(b_{\mathcal{T}},c_1,c_2)$. }
\begin{table}
\centering
\renewcommand*{\arraystretch}{1.5}
\begin{tabular}{|c|c|c|c|c|}
\hline
$Y_7$ & $\mathcal{N}$ & $\alpha$ & $B$ & $C$ \\
\hline\hline
$S^7/\mathbb{Z}_k$ (free) & $6$ & $\frac{\sqrt{2k}}{3}$ & $\frac{k}{24} - \frac{2}{3k}$ & $-\frac{3}{k}$ \\
\hline
$S^7/\mathbb{Z}_{N_f}$ (fixed points) & $4$ & $\frac{\sqrt{2N_f}}{3}$ & $-\frac{7N_f}{24} - \frac{1}{3N_f}$ & $-\frac{N_f}{2} -\frac{5}{2N_f}$ \\
\hline
$N^{0,1,0}/\mathbb{Z}_k$ & $3$ & $\frac{4\sqrt{k}}{3\sqrt{3}}$ & $-\frac{5k}{48} - \frac{1}{3k}$ & $-\frac{k}{4} - \frac{5}{4k}$ \\
\hline
$V^{5,2}/\mathbb{Z}_{N_f}$ & $2$ & $\frac{16\sqrt{N_f}}{27}$ & $-\frac{N_f}{6} - \frac{1}{4N_f}$ & $-\frac{9N_f}{16} - \frac{27}{16N_f}$ \\
\hline
$Q^{1,1,1}/\mathbb{Z}_{N_f}$ & $2$ & $\frac{4\sqrt{N_f}}{3\sqrt{3}}$ & $-\frac{N_f}{6}$ & $-\frac{N_f}{4} - \frac{3}{4N_f}$ \\
\hline
\end{tabular}
\caption{The localization data for various $\mathcal{N}\geq 2$ SCFTs. The first two entries are mirror dual of each other for $k=N_f=1$ and correspond to the ABJM and ADHM theories. The label in the second column indicates the amount of supersymmetry preserved by the theory for general choices of the integer parameters $(k,N_f)$. \label{tab:loc}}
\end{table}
We can use \eqref{eq:grav-N} to translate \eqref{eq:thermal-grav} into CFT quantities. Doing so, we obtain the leading behavior in the large $N$ limit with the expected $N^{3/2}$ scaling for M2-branes, as well as the first subleading correction: 
\begin{equation}\label{eq:bTc1c2M2}
\begin{split}
b_{\mathcal{T}} =&\; -\frac{8\pi^2}{9}\,A\,N^{3/2} - \frac{8\pi^2}{9}\,(a - v_1 + v_2)\,N^{1/2} \, , \\
c_1 =&\; \frac16\,A\,N^{3/2} + \frac{1}{12}\,(2a + v_1 + 2v_2)\,N^{1/2} \, , \\
c_2 =&\; -\frac1{16}\,A\,N^{3/2} - \frac{1}{16}\,(a - v_1 + v_2)\,N^{1/2} \, ,
\end{split}
\end{equation}
where we display $b_\mathcal{T}$ rather than $f_\mathcal{T}$ by using the relation \eqref{eq:thermal-rel}. Finally, using the holographic dictionary \eqref{eq:holo-dict} and Table~\ref{tab:loc}, we obtain the expression for the various thermal quantities purely in terms of CFT data. The results for various $\mathcal{N}\geq2$ SCFTs are given in Tables~\ref{tab:bT-sub} and~\ref{tab:ci-sub}. We emphasize that these results can be combined with \eqref{eq:g1-g2holo4der} to obtain the Casimir energy of these 3d $\mathcal{N}=2$ SCFTs to subleading order in the large $N$ limit. Moreover, the results for $c_{1,2}$ can be used to obtain the subleading $1/N$ corrections to the high-temperature density of states in the 3d SCFT as derived in \cite{Benjamin:2023qsc}.

We note that in \cite{Gubser:1998nz} the authors computed the $N^{3/2}$ and $N^{1/2}$ terms in the thermal free energy of the low-energy theory of M2-branes probing $\mathbb{C}^4$. This should correspond to the case $k=1$ in the ABJM theory we consider here. They find that the ratio of the coefficients of the $N^{1/2}$ term to that of the $N^{3/2}$ term is $\frac{128}{3} \times 2^{2/3} \times \pi^{16/3}$ (see (67) and (68) in \cite{Gubser:1998nz}) whereas from our result for $f_\mathcal{T}$ we find that this ratio is $\frac{15}{16}$. We believe that our result is correct and the discrepancy is due to the fact that in  \cite{Gubser:1998nz} the authors do not take into account the full set of higher-derivative corrections in 11d supergravity that affect the calculation of the $N^{1/2}$ term in the thermal free energy.\footnote{Our result for the $N^{3/2}$ term in the free energy is the same as the one in \cite{Gubser:1998nz}.} In contrast, the 4d higher-derivative method we have employed here automatically takes into account all leading higher-derivative corrections as evidenced by the non-trivial consistency checks between holography and supersymmetric localization performed in \cite{Bobev:2020egg,Bobev:2021oku,Bobev:2022jte,Bobev:2022eus,Bobev:2023lkx}.

\begin{table}[h]
\centering
\renewcommand*{\arraystretch}{1.8}
\begin{tabular}{|c|c|}
\hline
$Y_7$ & $b_\mathcal{T}$ \\
\hline\hline
$S^7/\mathbb{Z}_k$ (free) & $-\frac{8\pi^2\sqrt{2k}}{27}\,N^{3/2} + \frac{\pi^2(k^2 - 16)}{27\sqrt{2k}}\,N^{1/2}$ \\
\hline
$S^7/\mathbb{Z}_{N_f}$ (fixed points) & $-\frac{8\pi^2\sqrt{2N_f}}{27}\,N^{3/2} - \frac{\pi^2(7N_f^2 + 8)}{27\sqrt{2N_f}}\,N^{1/2}$ \\
\hline
$N^{0,1,0}/\mathbb{Z}_k$ & $-\frac{32\pi^2\sqrt{k}}{27\sqrt{3}}\,N^{3/2} - \frac{\pi^2(5k^2 + 16)}{27\sqrt{3k}}\,N^{1/2}$ \\
\hline
$V^{5,2}/\mathbb{Z}_{N_f}$ & $-\frac{128\pi^2\sqrt{N_f}}{243}\,N^{3/2} - \frac{16\pi^2(2N_f^2 + 3)}{243\sqrt{N_f}}\,N^{1/2}$ \\
\hline
$Q^{1,1,1}/\mathbb{Z}_{N_f}$ & $-\frac{32\pi^2\sqrt{N_f}}{27\sqrt{3}}\,N^{3/2} - \frac{8\pi^2N_f^2}{27\sqrt{3N_f}}\,N^{1/2}$ \\
\hline
\end{tabular}
\caption{The coefficient of the stress tensor one-point function \eqref{eq:bT-def} to subleading order in the large $N$ limit for various 3d $\mathcal{N}\geq 2$ M2-brane SCFTs placed on $S^1_\beta \times \mathbb{R}^2$. \label{tab:bT-sub}}
\end{table}
\begin{table}[h]
\centering
\renewcommand*{\arraystretch}{1.8}
\begin{tabular}{|c|c|c|}
\hline
$Y_7$ & $c_1$ & $c_2$ \\
\hline\hline
$S^7/\mathbb{Z}_k$ (free) & $\frac{\sqrt{k}}{9\sqrt{2}}\,N^{3/2} - \frac{k^2 + 20}{144\sqrt{2k}}\,N^{1/2}$ & $-\frac{\sqrt{k}}{24\sqrt{2}}\,N^{3/2} + \frac{k^2 - 16}{384\sqrt{2k}}\,N^{1/2}$ \\
\hline
$S^7/\mathbb{Z}_{N_f}$ (fixed points) & $\frac{\sqrt{N_f}}{9\sqrt{2}}\,N^{3/2} + \frac{N_f^2 - 22}{144\sqrt{2N_f}}\,N^{1/2}$ & $-\frac{\sqrt{N_f}}{24\sqrt{2}}\,N^{3/2} - \frac{7N_f^2 + 8}{384\sqrt{2N_f}}\,N^{1/2}$ \\
\hline
$N^{0,1,0}/\mathbb{Z}_k$ & $\frac{2\sqrt{k}}{9\sqrt{3}}\,N^{3/2} - \frac{k^2 + 14}{144\sqrt{3k}}\,N^{1/2}$ & $-\frac{\sqrt{k}}{12\sqrt{3}}\,N^{3/2} - \frac{5k^2 + 16}{384\sqrt{3k}}\,N^{1/2}$\\
\hline
$V^{5,2}/\mathbb{Z}_{N_f}$ & $\frac{8\sqrt{N_f}}{81}\,N^{3/2} - \frac{11N_f^2 + 57}{648\sqrt{N_f}}\,N^{1/2}$ & $-\frac{\sqrt{N_f}}{27}\,N^{3/2} - \frac{2N_f^2 + 3}{216\sqrt{N_f}}\,N^{1/2}$ \\
\hline
$Q^{1,1,1}/\mathbb{Z}_{N_f}$ & $\frac{2\sqrt{N_f}}{9\sqrt{3}}\,N^{3/2} + \frac{N_f^2 - 9}{72\sqrt{3N_f}}\,N^{1/2}$ & $-\frac{\sqrt{N_f}}{12\sqrt{3}}\,N^{3/2} - \frac{N_f^2}{48\sqrt{3N_f}}\,N^{1/2}$\\
\hline
\end{tabular}
\caption{The coefficients of the thermal effective action \eqref{eq:S-th} to subleading order in the large $N$ limit for various 3d $\mathcal{N}\geq 2$ M2-brane SCFTs placed on $S^1_\beta \times S^2$. \label{tab:ci-sub}}
\end{table}

\section{Comments on the $S^1 \times S^2$ free energy}
\label{sec:Sthermal}

In this section we make some general comments regarding the structure of the $S^1 \times S^2$ free energy in \eqref{eq:Sthsaddle} for Euclidean 3d CFTs. As already reviewed, the first two terms in a large temperature expansion are given by \eqref{eq:S-th}, see~\cite{Benjamin:2023qsc,Bhattacharyya:2007vs}. It is also instructive to collect further subleading terms, which can be done in specific CFTs using holography or the exact form of the grand canonical partition function. We proceed with illustrating this on three examples and based on these results, we formulate a general conjecture about the form of the $S^1 \times S^2$ free energy to all orders in the large $T$ expansion. \\

We can derive the $S^1 \times S^2$ free energy of holographic CFTs by studying the AdS-Kerr on-shell action along the lines of Section~\ref{sec:AdS-Kerr}. In particular, we find the following structure for the large temperature expansion:
\begin{equation}
\label{eq:I-Kerr-expand}
I_{2\partial}^{\text{Kerr}} = -\frac{16\pi^3 L^2}{27 G_N}\frac{(\beta/L)^{-2}}{1 + L^2\omega^2}\left[1 + \sum_{n\geq1}\,d_n\,\left(\frac{\beta}{\pi L}\right)^{2n}\,\sum_{m=0}^{n} c_{n,m}\,(L\,\omega)^{2m}\right] \, ,
\end{equation}	
where $\beta$ and $\omega$ are given in terms of the black hole parameters in \eqref{eq:beta-Kerr} and \eqref{eq:omega-Kerr}. The value of the coefficients $d_n$ and $c_{n,m}$ is given in Table \ref{tab:d-Kerr} and Table \ref{tab:c-Kerr}, respectively, for the first few values of $n$. It is straightforward to obtain higher order terms by making use of \eqref{eq:I-Kerr-2der} and \eqref{eq:r+-beta:Omega}. Note that we choose to split the expansion coefficients into $d_n$ and $c_{n,m}$ by demanding that $c_{n,m} \in \mathbb{Z}$ for all $(n,m)$ and that gcd$(\{c_{n_0,m}\}) = 1$ for all fixed $n_0$. This is a purely aesthetic choice.\\

\begin{table}[h]
\centering
\renewcommand*{\arraystretch}{1.5}
\begin{tabular}{|c||c|c|c|c|c|c|c|c|}
\hline
$n$ & 1 & 2 & 3 & 4 & 5 & 6 & 7 & 8 \\
\hline
$d_n$ & $-\frac{9}{16}$ & $\frac{27}{256}$ & $\frac{27}{4096}$ & $\frac{243}{65536}$ & $\frac{729}{1048576}$ & $\frac{729}{16777216}$ & $\frac{19683}{134217728}$ & $\frac{59049}{4294967296}$ \\
\hline
\end{tabular}
\caption{The coefficients $d_n$ appearing in the large temperature expansion of the two-derivative regularized on-shell action~\eqref{eq:I-Kerr-expand} of the Kerr-AdS black hole.\label{tab:d-Kerr}}
\end{table}
\begin{table}\footnotesize
\centering
\renewcommand*{\arraystretch}{1.3}
\begin{tabular}{|c||c|c|c|c|c|c|c|c|c|}
\hline
\diagbox[width=\dimexpr \textwidth/16+2\tabcolsep\relax, height=1cm]{$n$}{$m$} & 0 & 1 & 2 & 3 & 4 & 5 & 6 & 7 & 8 \\
\hline
$1$ & 2 & 1 & - & - & - & - & - & - & - \\
\hline
$2$ & 1 & 1 & 1 & - & - & - & - & - & - \\
\hline
$3$ & 2 & 3 & $-3$ & $-2$ & - & - & - & - & - \\
\hline
$4$ & 1 & 2 & 3 & 2 & 1 & - & - & - & - \\
\hline
$5$ & 2 & 5 & 2 & $-2$ & $-5$ & $-2$ & - & - & - \\
\hline
$6$ & 14 & 42 & 57 & 44 & 57 & 42 & 14 & - & - \\
\hline
$7$ & 2 & 7 & 9 & 5 & $-5$ & $-9$ & $-7$ & $-2$ & - \\
\hline
$8$ & 11 & 44 & 74 & 68 & 65 & 68 & 74 & 44 & 11 \\ 
\hline
\end{tabular}
\caption{The coefficients $c_{n,m}$ appearing in the large temperature expansion of the two-derivative regularized on-shell action~\eqref{eq:I-Kerr-expand} of the Kerr-AdS black hole.\label{tab:c-Kerr}}
\end{table}

For free CFTs, the grand canonical partition function can be computed exactly as discussed in Appendix C of~\cite{Benjamin:2023qsc}. In particular, for a free real scalar we have 
\begin{equation}
Z^s(T,\Omega) = \prod_{m_0\geq0}\prod_{m_1\in\mathbb{Z}}\frac{1}{1-e^{-\beta(2m_0 + |m_1| + \frac12 + \mathrm{i}m_1\Omega)}}\times\prod_{m_0\geq0}\prod_{m_1\in\mathbb{Z}}\frac{1}{1-e^{-\beta(2m_0 + |m_1| + \frac32 + \mathrm{i}m_1\Omega)}} \, ,
\end{equation}
whose logarithm can be expanded at large temperature as follows. First, we define
\begin{equation}
F_\alpha := \sum_{m_0\geq0}\sum_{m_1\in\mathbb{Z}}\sum_{k\geq1}\frac1{k}\,e^{-\beta k\bigl(2m_0 + \alpha + |m_1|(1 + \mathrm{i}s_1\Omega)\bigr)} \, ,
\end{equation}
where $s_1 = \text{sign}(m_1)$. With this definition, we have $\log Z^s = F_{1/2} + F_{3/2}$. The sums over $m_1$ and $m_0$ are geometric series which can be performed to find
\begin{equation}
F_\alpha = \sum_{k\geq 1}\frac{1}{k}\,\frac{e^{\beta k(2-\alpha)}}{1 + e^{2\beta k} - 2 e^{\beta k}\cos(\beta k \Omega)} \, .
\end{equation}
Each of the two contributions in $\log Z^s = F_{1/2} + F_{3/2}$ can be expanded at large temperature, and the sums over $k$ can be performed using $\zeta$-function regularization. The final result reads
\begin{equation}
\label{eq:logZb}
\log Z^s(T,\Omega) = \frac{\beta^{-2}}{1+\Omega^2}\Bigl[2\zeta(3) + \frac{1 + 2\Omega^2}{12}\zeta(1)\beta^2 + \sum_{n\geq 2} d^s_n\,\beta^{2n}\sum_{m=0}^n c^s_{n,m}\Omega^{2m}\Bigr] \, .
\end{equation}
We see from the leading term and \eqref{eq:S-th} that
\begin{equation}
f_\mathcal{T}^{s} = -\frac{\zeta(3)}{2\pi} \, ,
\end{equation}
which matches the result \eqref{eq:bT-free} for a free scalar field in 3d upon using the relation \eqref{eq:f-b}. We also note the appearance of a divergent $\zeta(1)\beta^0$ term which signals the presence of a $\log\beta$ term. This term was derived in \cite{Melia:2020pzd} using a method that does not rely on zeta-function regularization (see also~\cite{Kang:2022orq}). Besides this logarithmic term, which is associated with the presence of gapless degrees of freedom after the KK reduction to 2d, the structure of the free scalar CFT free energy is remarkably similar to the expansion for holographic CFTs implied by \eqref{eq:I-Kerr-expand}. The first few coefficients in the expansion \eqref{eq:logZb} are given in Tables \ref{tab:d-b} and \ref{tab:c-b}, respectively.
\begin{table}
\centering
\renewcommand*{\arraystretch}{1.5}
\begin{tabular}{|c||c|c|c|c|c|c|}
\hline
$n$ & 2 & 3 & 4 & 5 & 6 & 7 \\
\hline
$d^s_n$ & $\frac{1}{34560}$ & $\frac{1}{58060800}$ & $\frac{1}{58525286400}$ & $\frac{1}{14714929152000}$ & $\frac{1}{176755728973824000}$ & $\frac{691}{1052821396578631680000}$ \\
\hline
\end{tabular}
\caption{The coefficients $d_n^s$ appearing in the large temperature expansion of the free energy~\eqref{eq:logZb} for the free scalar theory.\label{tab:d-b}}
\end{table}

A similar computation can be done for the 3d free fermion theory, whose grand canonical partition function reads, see~\cite{Benjamin:2023qsc}
\begin{equation}
Z^f(T,\Omega) = \prod_{m_0\geq0}\prod_{m_1\in\mathbb{Z}+\frac12}\Bigl[1 + e^{-\beta(m_0 + |m_1| + \frac12 + \mathrm{i}m_1\Omega)}\Bigr] \, .
\end{equation}
Using the same method as in the free scalar case, we obtain the following large temperature expansion:
\begin{equation}
\label{eq:logZf}
\log Z^f(T,\Omega) = \frac{\beta^{-2}}{1+\Omega^2}\Bigl[\frac32\zeta(3) + \sum_{n\geq 2} d^f_n\,\beta^{2n}\sum_{m=0}^n c^f_{n,m}\Omega^{2m}\Bigr] \, ,
\end{equation}
with the values of the expansion coefficients given in Tables \ref{tab:d-f} and \ref{tab:c-f}. Once again, we observe the same structure of the large temperature expansion as for holographic CFTs and the free scalar theory. Furthermore, inspecting Table~\ref{tab:c-b} and Table~\ref{tab:c-f}, it is immediate to observe the relation
\begin{equation}\label{eq:cmnbosonfermion}
(-1)^{n+1} c_{n,n-m}^f = c_{n,m}^s \, ,
\end{equation}
for any given $(n,m)$. It will be interesting to understand if this relation is a mere coincidence or if there is a deeper reason for its validity.
\begin{table}[h]
\centering
\renewcommand*{\arraystretch}{1.5}
\begin{tabular}{|c||c|c|c|c|c|c|}
\hline
$n$ & 2 & 3 & 4 & 5 & 6 & 7 \\
\hline
$d^f_n$ & $\frac{1}{11520}$ & $\frac{1}{3870720}$ & $\frac{1}{928972800}$ & $\frac{17}{980995276800}$ & $\frac{31}{5356234211328000}$ & $\frac{691}{257099242143744000}$ \\
\hline
\end{tabular}
\caption{The coefficients $d_n^f$ appearing in the large temperature expansion of the free energy~\eqref{eq:logZf} for the free fermion theory.\label{tab:d-f}}
\end{table}

In view of the above results, it is tempting to conjecture that any 3d CFT admits the following large temperature expansion of its $S^1 \times S^2$ free energy: 
\begin{equation}
\label{eq:S-th-gen}
F_{S^1 \times S^2}(T,\Omega) = \frac{4\pi}{1+\Omega^2}\Bigl[f_{\mathcal{T}}\,T^2 + 2\,c_1\,(1 + \Omega^2) + \frac83\,c_2\,\Omega^2 + \sum_{n=1}^{\infty}\sum_{m=0}^{n}\mathfrak{c}_{m,n}\Omega^{2m}T^{-2n}\Bigr] \, .
\end{equation}
The coefficients $\mathfrak{c}_{m,n}$ depend on the CFT in question and can be determined explicitly and systematically for holographic CFTs as well as for free scalars and fermions, as shown above. We note that the results in \eqref{eq:S-th-gen} should be viewed as a perturbative expansion at large $T$ which presumably could receive exponentially small corrections. Indeed, for free scalars and fermions such non-perturbative corrections are present, see \cite{Benjamin:2023qsc}. We stress that in \eqref{eq:S-th-gen} we have not included the potential contribution of gapless modes to the $S^1 \times S^2$ free energy which could be present for some CFTs, as we saw for the free scalar field. These contributions can be associated to anomalies and the corresponding non-local terms in the thermal effective action for the metric and KK gauge field, see \cite{Bhattacharyya:2007vs,Benjamin:2023qsc} for a more detailed discussion of such terms. While we believe that the expansion in \eqref{eq:S-th-gen} is true for general CFTs it will be nice to establish its validity more rigorously using the methods of \cite{Benjamin:2023qsc}. It will also be desirable to understand whether it is possible to relate the coefficients $\mathfrak{c}_{m,n}$ to local data of the CFT and how to compute them for general interacting CFTs or to constrain their range using ideas similar to the ones employed in the conformal bootstrap.

\begin{table}[h]\footnotesize
\renewcommand*{\arraystretch}{1.2}
\centerline{\begin{tabular}{|c||c|c|c|c|c|c|c|c|}
\hline
\diagbox[width=\dimexpr \textwidth/16+2\tabcolsep\relax, height=1cm]{$n$}{$m$} & 0 & 1 & 2 & 3 & 4 & 5 & 6 & 7 \\
\hline
\hline
$2$ & 21 & 4 & $-24$ & - & - & - & - & - \\
\hline
$3$ & 155 & $-302$ & $-328$ & $160$ & - & - & - & - \\
\hline
$4$ & 2667 & $-13768$ & $-176$ & 13952 & $-2688$ & - & - & - \\
\hline
$5$ & 22995 & $-220962$ & $248528$ & 250432 & $-221568$ & 23040 & - & - \\
\hline
$6$ & 15559247 & $ -238754908$ & $613166088$ &  565888 & $-613859712$ & 238902272 & $-15566848$ & - \\
\hline
$7$ & 11180715 & $ -250354262$ & $1093392328$ & $-922951200$ & $-923533440$ & 1093632512 & $-250390528$ & 11182080 \\
\hline
\end{tabular}}
\caption{The coefficients $c^s_{n,m}$ appearing in the large temperature expansion of the free energy~\eqref{eq:logZb} for the free scalar theory.\label{tab:c-b}}
\end{table}
\begin{table}[h]\footnotesize
\renewcommand*{\arraystretch}{1.2}
\centerline{\begin{tabular}{|c||c|c|c|c|c|c|c|c|}
\hline
\diagbox[width=\dimexpr \textwidth/16+2\tabcolsep\relax, height=1cm]{$n$}{$m$} & 0 & 1 & 2 & 3 & 4 & 5 & 6 & 7 \\
\hline
\hline
$2$ & 24 & $-4$ & $-21$ & - & - & - & - & - \\
\hline
$3$ & 160 & $-328$ & $-302$ & $155$ & - & - & - & - \\
\hline
$4$ & 2688 & $-13952$ & $176$ & 13768 & $-2667$ & - & - & - \\
\hline
$5$ & 23040 & $-221568$ & $250432$ & 248528 & $-220962$ & 22995 & - & - \\
\hline
$6$ & 15566848 & $-238902272$ & $613859712$ & $-565888$ & $-613166088$ & 238754908 & $-15559247$ & - \\
\hline
$7$ & 11182080 & $-250390528$ & $1093632512$ & $-923533440$ & $-922951200$ & 1093392328 & $-250354262$ & 11180715 \\
\hline
\end{tabular}}
\caption{The coefficients $c^f_{n,m}$ appearing in the large temperature expansion of the free energy~\eqref{eq:logZf} for the free fermion theory.\label{tab:c-f}}
\end{table}

\section{Discussion}
\label{sec:discussion}

We used recent results in higher-derivative supergravity, supersymmetric localization, and holography to show how to explicitly calculate certain thermal observables in strongly interacting 3d SCFTs arising from M2-branes to order $N^{1/2}$ in the large $N$ expansion. Our analysis leads to several natural questions for future work as well as a number of possible generalizations, some of which we discuss below.

The class $\mathcal{R}$ 3d $\mathcal{N}=2$ SCFTs arise on the worldvolume of $N$ M5-branes wrapping three-manifolds after an appropriate topological twist. One can use similar methods to the ones described above to combine supersymmetric localization results, 3d-3d duality, and higher-derivative supergravity to compute observables in these SCFTs beyond the leading order in the large $N$ limit, see \cite{Bobev:2020zov} for more details and further references. We can leverage these results in the current context to also compute thermal observables in the class $\mathcal{R}$ SCFTs. As opposed to theories arising from M2-branes, see \eqref{eq:grav-N}, the leading term in the large $N$ limit for class $\mathcal{R}$ SCFTs is of order $N^3$ and the subleading one is of order $N$.\footnote{We assume that the 3-manifold wrapped by the M5-branes is compact and hyperbolic. In more general setups there could also be an $N^2$ term in the large $N$ expansion.} We can then use the results of \cite{Bobev:2020zov} to find
\begin{equation}
A = \frac{\text{vol} M_3}{3\pi^2}\,, \qquad a+v_2 = - \frac{\text{vol} M_3}{4\pi^2}\,, \qquad v_1 = \frac{\text{vol} M_3}{12\pi^2}\,.
\end{equation}
After performing the same analysis that led to \eqref{eq:bTc1c2M2} we find the following values for the thermal observables to order $N$ in the large $N$ expansion
\begin{equation}\label{eq:bTc1c2M5}
\begin{split}
b_{\mathcal{T}} =&\; -\frac{8\pi^2}{9}\,A\,N^{3} - \frac{8\pi^2}{9}\,(a - v_1 + v_2)\,N\,, \\
c_1 =&\; \frac16\,A\,N^{3} + \frac{1}{12}\,(2a + v_1 + 2v_2)\,N \, , \\
c_2 =&\; -\frac1{16}\,A\,N^{3} - \frac{1}{16}\,(a - v_1 + v_2)\,N \, .
\end{split}
\end{equation}
These results provide new information about the thermal properties of class $\mathcal{R}$ 3d $\mathcal{N}=2$ SCFTs at large $N$.

The coefficient $C_\mathcal{T}$ of the energy-momentum tensor two-point function in any 3d $\mathcal{N}=2$ SCFT can be computed using supersymmetric localization by taking appropriate derivatives of the squashed $S^3$ partition function, see \cite{Pestun:2016zxk} for a review. For the large $N$ limit of the 3d SCFTs arising from M2-branes one can apply this result to find the following expression at order $N^{1/2}$, see \cite{Bobev:2020egg,Bobev:2021oku},
\begin{equation}\label{eq:CTHD}
C_\mathcal{T}= \frac{64}{\pi}\left[\,A\,N^{3/2} +\,(a - v_1 + v_2)\,N^{1/2}\right]\,,
\end{equation}
where $(A,a,v_1,v_2)$ are defined in \eqref{eq:grav-N}. Comparing \eqref{eq:CTHD} to the expression for $b_{\mathcal{T}}$ in \eqref{eq:bTc1c2M2} we find that for this class of SCFTs the following relation holds
\begin{equation}\label{eq:bTcTrel}
b_{\mathcal{T}} = -\frac{\pi^3}{72}\,C_\mathcal{T} \, .
\end{equation}
The fact that $f_{\mathcal{T}}$, or equivalently $b_{\mathcal{T}}$, should be related to $C_\mathcal{T}$ at leading order in the large $N$ limit of holographic CFTs has been emphasized in~\cite{Kovtun:2008kw}. In~\cite{Buchel:2009sk} it was noted that higher-derivative terms may in principle spoil this universality. For the class of M2-brane SCFTs considered here, we find that the universality persists also at the first subleading order in the large $N$ limit.\footnote{This is also true for the class $\mathcal{R}$ SCFTs discussed above.} Given that many large $N$ partition functions of 3d SCFTs contain a $\log N$ term and that no such term is present in the large $N$ expansion of $C_\mathcal{T}$ (cf.~\cite{Agmon:2017xes,Chester:2020jay}), one may naively conclude that the relation breaks down at order $\log N$. However, it was argued in~\cite{Hristov:2021zai} that the coefficient of the $\log N$ term for holographic SCFTs is proportional to the Euler number of the Euclidean 4d supergravity background used to compute the gravitational path integral. Using this result we conclude that there is no $\log N$ term in the large $N$ expansion of $f_{\mathcal{T}}$ (and therefore of $b_{\mathcal{T}}$) since the corresponding AdS$_4$ soliton solution in Section~\ref{subsec:AdSSoliton} has $\chi=0$. This argument leads to the conclusion that the relation \eqref{eq:bTcTrel} holds for the first three orders in the large $N$ expansion of holographic SCFTs arising from M2-branes. Given this result, it is clearly of great interest to understand at which order the relation~\eqref{eq:bTcTrel} breaks down and what is the physical reason for its validity at large $N$.\footnote{It will be interesting to extend this analysis to thermal observables in Chern-Simons theories with fundamental matter in \cite{Aharony:2012ns}.}
In general, we expect some large $N$ approximation to~\eqref{eq:bTcTrel} to hold only in holographic theories. To support this expectation, let us consider the large $N$ limit of the 3d $O(N)$ vector model as a prototypical large $N$ theory that do not admit an Einstein gravity holographic dual description. The stress tensor two-point function $C_\mathcal{T}$ was obtained to subleading order in $N$ in~\cite{Petkou:1995vu}, while the subleading term in the large $N$ expansion of the one-point coefficient $b_\mathcal{T}$ was derived in~\cite{Diatlyk:2023msc}. Using these results, it is easy to check that the ratio of the two quantities depends on $N$ and is given by $C_\mathcal{T}/b_\mathcal{T} = -(72/\pi^3)(1 - 0.0321833 N^{-1})$ to first subleading order, spoiling the strict large $N$ universality advocated in \cite{DeWolfe:2019etx,Romatschke:2019ybu} for this model. A similar violation of~\eqref{eq:bTcTrel} due to higher-derivative corrections in Einsteinian cubic gravity has been observed in~\cite{Bueno:2018xqc}.

For the $S^1 \times S^2$ free energy on the other hand, one expects that there will be non-trivial $\log N$ corrections since the Euclidean AdS-Kerr solution has $\chi=2$. While not much is known about such logarithmic corrections in the context of AdS$_4$ holography it is reasonable to speculate that the coefficient of $\log N$ terms in the large $N$ expansion of physical observables should not depend on continuous parameters like temperature and chemical potentials. If this is indeed true we can use the general structure of the $S^1 \times S^2$ free energy in \eqref{eq:S-th-gen} above to conclude that only the coefficient $c_1$ can receive logarithmic corrections in the large $N$ limit of holographic CFTs. It will be very interesting to explore this further and establish the validity of our comments more rigorously. 

It will also be desirable to generalize the results discussed in Section~\ref{sec:Sthermal} to higher dimensions $d >3$.~The explicit calculations for free field theories should be relatively straightforward to perform. The same applies for the holographic analysis of the Kerr-AdS solution in higher dimensions. Based on these results it should be possible to conjecture a general form of the $S^1\times S^{d-1}$ thermal free energy analogous to~\eqref{eq:S-th-gen}. To this end it will also be nice to extend these results to the large temperature limit of thermal CFTs in the presence of chemical potentials for continuous global symmetries. A natural first line of attack to get access to such thermal partition functions is to analyze an appropriate limit of the regularized on-shell action of the AdS-Kerr-Newman charged black hole solution. This analysis could be facilitated in $d=4$ by following the steps outlined in this work and exploiting the recent results on higher-derivative 5d gauged supergravity and holography in \cite{Bobev:2021qxx,Bobev:2022bjm,Cassani:2022lrk,Cassani:2023vsa}.

The conjecture~\eqref{eq:S-th-gen} for the large temperature expansion of the $S^1\times S^{2}$ thermal free energy poses the question of why there are only even powers of $T$ and $\Omega$ present in the expansion. We believe that this is dictated by both the dimensionality of space, i.e. $d=3$, and the structure of the thermal effective action discussed in~\cite{Benjamin:2023qsc}. In particular, it is possible to check that with $d=3$ and considering only parity-preserving local terms in the thermal effective action, one can generate only even powers of $T$ and $\Omega$. It is desirable to make this analysis more rigorous and to explore also the corresponding structure of the $S^1\times S^{d-1}$ thermal free energy for other values of $d$ and in the presence of parity breaking terms.

There are classes of 3d $\mathcal{N}=2$ holographic SCFTs which arise from D-branes and thus, in addition to the large $N$ limit, also admit a weak/strong coupling expansion. It will be most interesting to understand how to study thermal observables of the type discussed above for these classes of models beyond the leading order terms in the large $N$ and strong (or weak) coupling limit. This will likely shed light on some of the questions we discussed above and will provide important insights into the properties of string theory on non-supersymmetric backgrounds.

\section*{Acknowledgments}

We are grateful to Nathan Benjamin, Anthony Charles, Kiril Hristov, Zohar Komargodski, Eric Perlmutter, Silviu Pufu, and Yifan Wang for useful discussions. This research is supported by the FWO projects G003523N and G094523N. NB and JH are also supported in part by the KU Leuven C1 grant ZKD1118 C16/16/005 and by Odysseus grant G0F9516N from the FWO. VR is supported by a public grant as part of the Investissement d'avenir project, reference ANR-11-LABX-0056-LMH, LabEx LMH. NB and VR are grateful to the ENS Paris for warm hospitality during part of this project. VR is partly supported by a Visibilité Scientifique Junior Fellowship from LabEx LMH and is grateful to the CCPP at New York University for hospitality during the final stages of this project.


\bibliography{ThermalHD}
\bibliographystyle{JHEP}

\end{document}